# Hydrodynamic Impulse in a Compressible Fluid

Bhimsen K. Shivamoggi
University of Central Florida
Orlando, FL 32816-1364

**Abstract**
A suitable expression for hydrodynamic impulse in a compressible fluid is deduced. The development of appropriate impulse formulation for compressible Euler equations confirms the propriety of the hydrodynamic impulse expression for a compressible fluid given here. Implications of the application of this formulation to a compressible vortex ring are pointed out. A variational characterization for an axisymmetric vortex system moving steadily in an ideal, compressible fluid is discussed.

# 1. Introduction

Impulse[1] formulations of Euler (and Navier-Stokes) equations were considered by Kuz'min [1] and Oseledets [2] and are suitable for the study of topological properties of vortex lines. Different impulse formulations are produced by various possible gauge transformations (Russo and Smereka [3]). In the Kuz'min-Oseledets gauge, the force impulse density, also called the impulse velocity **q** has an interesting geometric meaning: it describes the evolution of material surfaces; its direction is orthogonal to the material surface element, and its length is proportional to the area of the surface element. The extension of the Kuz'min-Oseledets formulation to the compressible barotropic case was considered in a brief way by Tur and Yanovsky [4]. However, in the latter formulation, the velocity **q** no longer has the physical significance of being the force impulse density. The purpose of this paper is to deduce first a suitable expression for hydrodynamic impulse in a compressible fluid. This turns out to be rather different from what might be expected from a straightforward modification of the corresponding expression for the incompressible case. The development of appropriate impulse formulation for compressible Euler equations restores for the velocity **q** the physical significance of being the force impulse density and confirms the propriety of the hydrodynamic impulse expression for a compressible fluid given here. Implications of the application of this formulation to a compressible vortex ring, and more generally, a variational characterization for an axisymmetric vortex system moving steadily in an ideal, compressible fluid are discussed.

# 2. Hydrodynamic Impulse in a Compressible Fluid

The Euler equations governing compressible flows are (in usual notation)

$$\frac{\partial \rho}{\partial t} + \nabla \cdot (\rho \mathbf{v}) = 0 \tag{1}$$

$$\frac{\partial \mathbf{v}}{\partial t} + (\mathbf{v} \cdot \nabla) \mathbf{v} = -\frac{1}{\rho} \nabla p. \tag{2}$$

Let us assume that $\partial \rho / \partial t = 0$, implying that the density is frozen in time. Equation (1) then implies the existence of a vector potential $\mathbf{A}(\mathbf{x}, t)$ such that (Shivamoggi and van Heijst [5])

$$\rho \mathbf{v} = \nabla \times \mathbf{A} \tag{3}$$

with the gauge condition,

$$\nabla \cdot \mathbf{A} = 0. \tag{4}$$

Using (4), we have from (3),

$$\nabla \times \rho \mathbf{v} = -\nabla^2 \mathbf{A} \tag{5}$$

---

[1]In view of the possible unbounded extent of the fluid, care needs to be exercised in relating it to the change in momentum from one instant to the next. This caveat has apparently led to the result that the impulse concept has not been exploited in fluid dynamics as fully as one might expect.

which has the solution,
$$\mathbf{A}(\mathbf{x}) = \frac{1}{4\pi} \int_{\mathcal{V}} \frac{\nabla' \times (\rho' \mathbf{v}')}{|\mathbf{x} - \mathbf{x}'|} d\mathbf{x}' \tag{6}$$
where we have suppressed the argument $t$ and primes refer to the argument $\mathbf{x}'$, and $\mathcal{V}$ is a material volume in the fluid enclosed by a surface $S$. Expanding
$$\frac{1}{|\mathbf{x} - \mathbf{x}'|} = \frac{1}{r} + \frac{\mathbf{x} \cdot \mathbf{x}'}{r^3} + O\left(\frac{1}{r^3}\right) \tag{7}$$
where $r \equiv |\mathbf{x}|$, we obtain from (6),
$$\mathbf{A}(\mathbf{x}) \sim \frac{1}{4\pi r} \int_{\mathcal{V}} (\nabla' \times \rho' \mathbf{v}') \, d\mathbf{x}' + \frac{1}{4\pi r^3} \int_{\mathcal{V}} (\mathbf{x} \cdot \mathbf{x}') (\nabla' \times \rho' \mathbf{v}') \, d\mathbf{x}' + O\left(\frac{1}{r^3}\right). \tag{8}$$
On using appropriate boundary conditions on $S$, we have
$$\begin{aligned}\int_{\mathcal{V}} (\nabla' \times \rho \mathbf{v}')_i \, d\mathbf{x}' &= \int_{\mathcal{V}} \frac{\partial}{\partial x'_j} \left[ x'_i (\nabla' \times \rho' \mathbf{v}')_j \right] d\mathbf{x}' \\ &= \oint_S x'_i (\nabla' \times \rho' \mathbf{v}') \cdot \hat{\mathbf{n}} \, dS' = 0. \end{aligned} \tag{9}$$

Next, note
$$(\mathbf{x} \cdot \mathbf{x}') (\nabla' \times \rho' \mathbf{v}') = \mathbf{x} \times [(\nabla' \times \rho' \mathbf{v}') \times \mathbf{x}'] + [(\nabla' \times \rho' \mathbf{v}') \cdot \mathbf{x}] \mathbf{x}' \tag{10}$$
and the result (Saffman [6]),
$$\int_{\mathcal{V}} [(\nabla' \times \rho' \mathbf{v}') \cdot \mathbf{x}] \mathbf{x}' d\mathbf{x}' = -\int_{\mathcal{V}} (\mathbf{x}' \cdot \mathbf{x}) (\nabla' \times \rho' \mathbf{v}') \, d\mathbf{x}'. \tag{11}$$
Using (9) - (11), (8) becomes
$$\mathbf{A}(\mathbf{x}) \sim \frac{1}{8\pi r^3} \mathbf{x} \times \int_{\mathcal{V}} (\nabla' \times \rho' \mathbf{v}') \times \mathbf{x}' d\mathbf{x}'. \tag{12}$$

Noting that, for a dipole of strength $\mathbf{M}$ at the origin, we have
$$\mathbf{A} = \frac{\mathbf{M} \times \mathbf{x}}{r^3} \tag{13}$$
we obtain, on comparing (12) and (13),
$$\mathbf{M} = \frac{1}{4\pi} \mathbf{I} \tag{14}$$
where $\mathbf{I}$ is the hydrodynamic impulse[2] for a compressible fluid,[3]
$$\mathbf{I} \equiv \frac{1}{2} \int_{\mathcal{V}} \mathbf{x} \times (\nabla \times \rho \mathbf{v}) \, d\mathbf{x}. \tag{15}$$

---

[2] The hydrodynamic impulse can be viewed as the total mechanical impulse of the non-conservative body force that is applied to a limited fluid volume to generate instantaneously from rest the given motion of the whole of the fluid at any time t (Lamb [7]).

[3] It may be mentioned that (15) was considered by Lele, as mentioned in an abstract [8], as a possible extension of hydrodynamic impulse for a compressible fluid.

Observe that (15) is rather different from what might be expected from a straightforward modification of the corresponding expression for the incompressible case, which is (Batchelor [9])

$$\mathbf{I} = \frac{\rho}{2} \int_{\mathcal{V}} \mathbf{x} \times \boldsymbol{\omega} \, d\mathbf{x} \tag{16}$$

where $\boldsymbol{\omega}$ is the vorticity,

$$\boldsymbol{\omega} \equiv \nabla \times \mathbf{v}. \tag{17}$$

One of the attractive properties of hydrodynamic impulse $\mathbf{I}$ is that it is an invariant of the motion in an unbounded incompressible fluid (Batchelor [9]). This result can be shown to continue to hold even for a compressible fluid. In order to see this, first note that equations (1) and (2) may be combined to give

$$\frac{\partial}{\partial t}(\rho \mathbf{v}) + \nabla \cdot (\rho \mathbf{v} \mathbf{v}) = -\nabla p. \tag{18}$$

We then have

$$\begin{aligned}
\frac{d\mathbf{I}}{dt} &= \frac{1}{2} \int_{\mathcal{V}} \mathbf{x} \times \left[ \nabla \times \frac{\partial}{\partial t}(\rho \mathbf{v}) \right] d\mathbf{x} \\
&= \int_{\mathcal{V}} \frac{\partial}{\partial t}(\rho \mathbf{v}) \, d\mathbf{x} \\
&= -\int_{\mathcal{V}} [\nabla p + \nabla \cdot (\rho \mathbf{v} \mathbf{v})] \, d\mathbf{x} \\
&= \oint_{S} (p \hat{\mathbf{n}} + \rho \mathbf{v} \mathbf{v} \cdot \hat{\mathbf{n}}) \, dS \\
&= 0
\end{aligned} \tag{19}$$

on using appropriate boundary conditions on $S$.

## 3. Impulse Formulations of Compressible Euler Equations

We now generalize Kuz'min's [1] impulse formulation to compressible Euler equations. Consider a change in the gauge of the fluid velocity field given by[4]

$$\mathbf{q} = \mathbf{v} + \frac{1}{\rho} \nabla \phi. \tag{20}$$

$\phi$ is an arbitrary scalar field and signifies a gauge freedom which is exploited in Section 4. (20) constitutes a generalization of Weber's transform (Lamb [7]) to a compressible fluid (Kuznetsov [11]). Observe the vorticity mismatch between $\mathbf{q}$ and $\mathbf{v}$ - $\mathbf{q}$ does not have the same vorticity as $\mathbf{v}$, i.e., $\nabla \times \mathbf{q} \neq \nabla \times \mathbf{v}$. This anomaly is more than offset by the fact that $\mathbf{q}$ is the force impulse density for the compressible case (see below).

---

[4]It may be mentioned that (20) was briefly considered by Buttke [10] as a generalization of Kuz'min's [1] formulation for a compressible fluid, but the correctness of (20) and the significance of $\mathbf{q}$ was not apparent thereof because the appropriate expression for the hydrodynamic impulse in a compressible fluid, namely (15), was not at hand.

Using (20), (15) becomes

$$
\begin{aligned}
\mathbf{I} &= \frac{1}{2} \int_{\mathcal{V}} \mathbf{x} \times (\nabla \times \rho \mathbf{q}) \, d\mathbf{x} \\
&= \frac{1}{2} \int_{\mathcal{V}} \rho \mathbf{x} \times (\nabla \times \mathbf{q}) \, d\mathbf{x} + \frac{1}{2} \int_{\mathcal{V}} \mathbf{x} \times (\nabla \rho \times \mathbf{q}) \, d\mathbf{x} \\
&= \int_{\mathcal{V}} \rho \mathbf{q} \, d\mathbf{x} + \frac{1}{2} \int_{\mathcal{V}} \mathbf{x} \times (\mathbf{q} \times \nabla \rho) \, d\mathbf{x} + \frac{1}{2} \int_{\mathcal{V}} \mathbf{x} \times (\nabla \rho \times \mathbf{q}) \, d\mathbf{x} \\
&= \int_{\mathcal{V}} \rho \mathbf{q} \, d\mathbf{x}.
\end{aligned}
\quad (21)
$$

So, $\mathbf{q}$ is indeed the force impulse density for a compressible fluid. This provides the rationale for introducing $\mathbf{q}$ as in (20).

## 4. Potential Helicity Like Lagrange Invariant

Lagrange invariants for compressible Euler equations are in general very difficult, if not impossible, to find. Rather drastic restrictions on the compressibility condition become necessary, as is evident in the following.

Using (20), equation (18) becomes

$$
\rho \left[ \frac{\partial \mathbf{q}}{\partial t} + (\mathbf{v} \cdot \nabla) \mathbf{q} \right] - \frac{\partial}{\partial t} (\nabla \phi) - \nabla \cdot (\mathbf{v} \nabla \phi) = -\nabla p \quad (22a)
$$

which may be rewritten as

$$
\rho \left( \frac{\partial q_i}{\partial t} + v_j \frac{\partial q_i}{\partial x_j} \right) - \frac{\partial}{\partial t} \left( \frac{\partial \phi}{\partial x_i} \right) - \frac{\partial}{\partial x_j} \left( v_j \frac{\partial \phi}{\partial x_i} \right) = -\frac{\partial p}{\partial x_i}. \quad (22b)
$$

Note, on using (20), we have

$$
\begin{aligned}
\frac{\partial}{\partial x_j} \left( v_j \frac{\partial \phi}{\partial x_i} \right) &= -v_j \frac{\partial^2 \phi}{\partial x_j \partial x_i} - \frac{\partial v_j}{\partial x_j} \frac{\partial \phi}{\partial x_i} \\
&= -\frac{\partial}{\partial x_i} \left( v_j \frac{\partial \phi}{\partial x_j} \right) + \frac{\partial v_j}{\partial x_i} \frac{\partial \phi}{\partial x_j} - \frac{\partial v_j}{\partial x_j} \frac{\partial \phi}{\partial x_i} \\
&= -\frac{\partial}{\partial x_i} \left( v_j \frac{\partial \phi}{\partial x_j} \right) + \frac{\partial v_j}{\partial x_i} (\rho q_j - \rho v_j) - \frac{\partial v_j}{\partial x_j} \frac{\partial \phi}{\partial x_i}.
\end{aligned}
\quad (23)
$$

Next, we assume a barotropic case, namely,

$$
p = p(\rho) \quad (24a)
$$

so we may introduce[5]

$$
P(\mathbf{x}) \equiv \int \frac{dp}{\rho}. \quad (24b)
$$

---

[5]Barotropic behavior typically results on assuming that either the specific entropy or the temperature is constant in space and time - P then represents the specific enthalpy.

Using (23) and (24), equation (22b) becomes

$$\frac{\partial q_i}{\partial t} + v_j \frac{\partial q_i}{\partial x_j} = -q_j \frac{\partial v_j}{\partial x_i} - \frac{\partial}{\partial x_i}\left(P - \frac{\partial \phi}{\partial t} - \frac{v_j}{\rho}\frac{\partial \phi}{\partial x_j} - \frac{v_j^2}{2}\right) - \frac{v_j}{\rho^2}\left(\frac{\partial \rho}{\partial x_j}\frac{\partial \phi}{\partial x_i} - \frac{\partial \rho}{\partial x_i}\frac{\partial \phi}{\partial x_j}\right). \quad (25)$$

Equation (25) shows that we need to impose the following gauge condition,

$$\frac{\partial \phi}{\partial t} + \frac{1}{\rho}(\mathbf{v}\cdot\nabla)\phi + \frac{1}{2}\mathbf{v}^2 - P = 0 \quad (26)$$

as well as assume a restricted compressibility condition, given by

$$\rho = \rho(\phi). \quad (27)$$

Equation (25) then becomes

$$\frac{\partial \mathbf{q}}{\partial t} + (\mathbf{v}\cdot\nabla)\mathbf{q} = -(\nabla\mathbf{v})^T \mathbf{q}. \quad (28)$$

On the other hand, taking the curl of equation (2) and using equations (1) and (24), we obtain the following equation for the potential vorticity $\boldsymbol{\omega}/\rho$,

$$\frac{\partial}{\partial t}\left(\frac{\boldsymbol{\omega}}{\rho}\right) + (\mathbf{v}\cdot\nabla)\left(\frac{\boldsymbol{\omega}}{\rho}\right) = \left(\frac{\boldsymbol{\omega}}{\rho}\cdot\nabla\right)\mathbf{v}. \quad (29)$$

On combining equation (28) with equation (29), we obtain

$$\left[\frac{\partial}{\partial t} + (\mathbf{v}\cdot\nabla)\right]\left(\frac{\mathbf{q}\cdot\boldsymbol{\omega}}{\rho}\right) = 0 \quad (30)$$

which leads to the following potential helicity like Lagrange invariant for the restricted compressible barotropic flow,

$$\frac{\mathbf{q}\cdot\boldsymbol{\omega}}{\rho} = const. \quad (31)$$

(31) is not exactly the potential helicity because $\boldsymbol{\omega} \neq \nabla \times \mathbf{q}$. Nonetheless, (31) admits a simple physical interpretation, as follows.

If $\mathbf{l}$ is a vector field associated with an infinitesimal line element of the fluid, $\mathbf{l}$ evolves according to (Batchelor [9]),

$$\left[\frac{\partial}{\partial t} + (\mathbf{v}\cdot\nabla)\right]\mathbf{l} = (\mathbf{l}\cdot\nabla)\mathbf{v} \quad (32)$$

which is identical to the potential vorticity equation (29). Therefore, the potential vortex lines evolve as fluid line elements.

On the other hand, if $\mathbf{S}$ is a vector field associated with an oriented material surface element of the fluid, $\mathbf{S}$ evolves according to (Batchelor [9]),

$$\left[\frac{\partial}{\partial t} + (\mathbf{v}\cdot\nabla)\right](\rho\mathbf{S}) = -(\nabla\mathbf{v})^T(\rho\mathbf{S}) \quad (33)$$

which is identical to the equation of evolution of the force impulse density $\mathbf{q}$, namely, equation (28a). Therefore, the field lines of the force impulse density $\mathbf{q}$ evolve as oriented fluid surface mass elements - the direction of **q** is orthogonal to the surface mass element $\rho\mathbf{S}$ and the length of **q** is proportional to the area of the surface mass element $\rho\mathbf{S}$.

Thus, the potential helicity like Lagrange invariant (31) is physically equivalent to the mass conservation of the fluid element.

## 5. Beltrami States for Compressible Flows

It is well known that a significant class of exact solutions of the fluid dynamics equations emerges under the Beltrami condition - the local vorticity is proportional to the stream function. These Beltrami solutions are also known to correlate well with real fluid behavior - one example is the Larichev-Reznik [12] nonlinear dipole-vortex localized structure; another example is a laminar vortex ring for which the vorticity contours near the center of the core, become more elliptical and nearly parallel to the stream function contours (Stanaway et al. [13]), (see Section 6 for further discussion on vortex rings).

We have for the compressible Beltrami state (Shivamoggi and van Heijst [5]),

$$\boldsymbol{\omega} = a\rho\mathbf{v} \tag{34}$$

where $a$ is an arbitrary constant.

On the other hand, noting that the force impulse density **q** evolves like the surface mass element $\rho\mathbf{S}$, (31) becomes

$$(\rho\mathbf{S})\cdot\mathbf{v} = const \tag{35}$$

which physically signifies the constancy of mass flux - this is of course inevitable because equation (34) corresponds to a steady state.

## 6. Application to a Compressible Vortex Ring

It is well known that an isolated vortex ring (see Shariff and Leonard [14] for a review) in an unbounded ideal fluid will move with a constant velocity in a direction perpendicular to its plane (Kelvin [15])[6]. In an incompressible fluid, the volume of a vortex ring must remain constant so that the vorticity will be proportional to the length of the ring $2\pi R$, R being the toroidal radius. This leads to the result that vortex rings will expand and contract about the axis of symmetry during their motion with the fluid. The hydrodynamic impulse provides a very convenient tool[7] to describe some important aspects of vortex ring motion (Saffman [20]).

Consider an axisymmetric thin-core compressible vortex ring of toroidal radius R, vorticity $\boldsymbol{\omega}$, density $\rho = \rho(r)$, r being the poloidal radius and core radius $r = a$. The hydrodynamic impulse for the ring is given from (15), on assuming the ring to have a small cross section,

---

[6]In a real fluid, a vortex ring slows down and grows to larger overall diameter during motion, as shown by the laboratory experiments of Maxworthy [16] - [18] which generated vortex rings by ejecting a short pulse of fluid from an orifice. The vorticity produced by viscous effects at the orifice exit rolls up to form a vortex ring.

[7]The impulse integral is superior to the momentum integral which turns out not to be absolutely convergent in an unbounded fluid, even for compact vortical structures like vortex rings (Buhler [19]).

by

$$\begin{aligned}\mathbf{I} &= \frac{1}{2}\int_{\mathcal{V}} \mathbf{x} \times (\nabla \times \rho \mathbf{v})\, d\mathbf{x} \\ &= \frac{1}{2}\int_{\mathcal{V}} \mathbf{x} \times (\rho\,\boldsymbol{\omega} + \nabla\rho \times \mathbf{v})\, d\mathbf{x} \\ &= (\rho + \rho' a)\,\pi R^2 \Gamma\,\hat{\mathbf{i}}_z = const\end{aligned} \quad (36)$$

where $\Gamma \equiv \omega \cdot \pi a^2$ is the circulation around the vortex ring, $\hat{\mathbf{i}}_z$ is a unit vector along the axis of the ring and prime here denotes differentiation with respect to the argument $r$.

In the incompressible limit $\rho = const$, (36) is in accord with the laboratory experimental results of Maxworthy [16] - [18] and numerical computations of Stanaway et al. [13] which showed a decrease in circulation of the ring due to vorticity loss via detrainment of vortical fluid into the wake of the ring and a concomitant increase in the radius of the ring[8]. On the other hand, in the compressible case, as indicated by (36), the vorticity loss to a wake, can be offset by the density gradient ($\rho' > 0$) in the core so the vortex ring will not have to expand as much as it would in the incompressible case.[9]

Benjamin [22] pointed out that an axisymmetric vortex ring moving steadily in an ideal, incompressible fluid has a variational characterization - it corresponds to maximizing the kinetic energy on an isovortical sheet subject to constant hydrodynamic impulse, the constant speed of propagation being the Lagrange multiplier[10]. It is pertinent to inquire if an axisymmetric vortex system moving steadily in an ideal, compressible fluid also has a variational characterization.

The total kinetic energy for a compressible fluid is given by

$$E = \frac{1}{2}\int_{\mathcal{V}} \rho \mathbf{v}^2 d\mathbf{x}. \quad (37)$$

On using (3), (37) may be rewritten as

$$E = \frac{1}{2}\int_{\mathcal{V}} \mathbf{A} \cdot \boldsymbol{\omega} d\mathbf{x}. \quad (38a)$$

which, for an axisymmetric vortex system, becomes

$$E = \frac{1}{2}\int_{\mathcal{V}} A_\theta\, \omega_\theta\, d\mathbf{x} \quad (38b)$$

$A_\theta$ being the Stokes stream function. On the other hand, the hydrodynamic impulse for an axisymmetric vortex system is given by

$$\mathbf{I} = I\hat{\mathbf{i}}_z,\ I = \int_{\mathcal{V}} r\left[\rho\,\omega_\theta + \left(\frac{\partial \rho}{\partial z}v_r - \frac{\partial \rho}{\partial r}v_z\right)\right] d\mathbf{x}. \quad (39)$$

Choosing $\boldsymbol{\omega}$ to be the canonical variable and maximizing E keeping I constant, we obtain

$$\frac{\delta E}{\delta \omega_\theta} = \lambda \frac{\delta I}{\delta \omega_\theta} \quad (40)$$

---

[8]Saffman [19] showed that a vortex ring exhibits similar trends subject to viscous diffusion.

[9]It may be mentioned that compressibility effects have been found also to slow down a vortex ring (Moore [21]).

[10]Alternatively, this may be effected by stipulating that the hydrodynamic impulse and position are conjugate canonical variables (Roberts and Donnelly [23]) in the sense that $\partial E/\partial I = v_z$, E being the kinetic energy.

which, on using (38) and (39), gives
$$\psi_\theta = \lambda \rho r. \tag{41}$$

(41) leads to
$$\frac{1}{r}\frac{\partial}{\partial r}(r\psi_\theta) = \rho v_z = 2\lambda\rho + \lambda r \frac{\partial \rho}{\partial r} \tag{42a}$$

from which,
$$\lambda = \frac{v_z/2}{[1 + (r/2\rho)(\partial\rho/\partial r)]}. \tag{42b}$$

Since the Lagrange multiplier $\lambda$ is a constant, the variational characterization (40) underlying a steadily moving axisymmetric vortex system in an ideal, compressible fluid then holds if the density varies radially according to
$$1 + \frac{r}{2\rho}\frac{\partial \rho}{\partial r} = const \tag{43a}$$

implying a linear radial dependence,
$$\rho \sim r. \tag{43b}$$

## 7. Discussion

In this paper, a suitable expression for hydrodynamic impulse in a compressible fluid is deduced. This turns out to be rather different from what might be expected from a straightforward modification of the corresponding expression for the incompressible case. The development of appropriate impulse formulation for compressible Euler equations confirms the propriety of the hydrodynamic impulse expression for a compressible fluid given here. Implications of the application of this formulation to a compressible vortex ring are pointed out. An axisymmetric vortex system moving steadily in an ideal, compressible fluid admits a variational characterization if the density has a linear radial dependence. Thanks to the density gradient in the core, a compressible vortex ring will not have to expand as much as it would in the incompressible case in order to offset the vorticity loss to a wake.

## Acknowledgments

The author is thankful to Dr. S. Kurien for helpful discussions on impulse formulations, Professor G. J. F. van Heijst for helpful discussions on vortex rings, and Dr. K. Shariff and Professors S. K. Lele, A. Leonard and K. Ohkitani for their helpful comments and suggestions.

## References

[1] G. A. Kuz'min: *Phys. Lett. A* **96**, 88, (1983).

[2] V. I. Oseledets: *Russ. Math. Surveys* **44**, 210, (1989).

[3] G. Russo and P. Smereka: *J. Fluid Mech.* **391**, 189, (1999).

[4] A. V. Tur and V. V. Yanovsky: *J. Fluid Mech.* **248**, 67, (1993).

[5] B. K. Shivamoggi and G. J. F. van Heijst: *Phys. Lett. A* **372**, 5688, (2008).

[6] P. G. Saffman: *Vortex Dynamics*, Cambridge Univ. Press, (1995).

[7] H. Lamb: *Hydrodynamics*, Cambridge Univ. Press, (1932).

[8] S. K. Lele: *Bull. Amer. Phys. Soc.* **35**, 2226, (1990).

[9] G. K. Batchelor: *An Introduction to Fluid Dynamics*, Cambridge Univ. Press, (1967).

[10] T. F. Buttke: in *Vortex Flows and Related Numerical Methods*, Eds. J. T. Beale, G. H. Cottet and S. Huberson, p. 39, Kluwer, (1993).

[11] E. A. Kuznetsov: *J. Fluid Mech.* **600**, 167, (2008).

[12] V. D. Larichev and G. M. Reznik: *Oceanology* **16**, 547, (1976).

[13] S. Stanaway, B. J. Cantwell and P. R. Spalart: NASA TM-101041, (1988).

[14] K. Shariff and A. Leonard: *Ann. Rev. Fluid Mech.* **24**, 235, (1992).

[15] Lord Kelvin: *London Edinburgh Dublin Phil. Mag. J. Sci.* Fifth Series **10**, 155, (1880).

[16] T. Maxworthy: *J. Fluid Mech.* **51**, 15, (1972).

[17] T. Maxworthy: *J. Fluid Mech.* **64**, 227, (1974).

[18] T. Maxworthy: *J. Fluid Mech.* **81**, 465, (1977).

[19] O. Buhler: *Ann. Rev. Fluid Mech.* **42**, 205, (2010).

[20] P. G. Saffman: *Studies Appl. Math.* **49**, 371, (1970).

[21] D. W. Moore: *Proc. Roy. Soc. (London)* **A 397**, 87, (1985).

[22] T. B. Benjamin: in *Lecture Notes in Mathematics*, Vol. **503**, Eds. P. Germain and B. Nayrolle, Springer Verlag, p. 8, (1976).

[23] P. H. Roberts and R. J. Donnelly: *Phys. Lett. A* **31**, 137, (1970).